\documentclass[journal,table]{IEEEtran}
\usepackage{adjustbox}
\usepackage[T1]{fontenc}
\usepackage{array}
\usepackage{afterpage}
\usepackage{amsmath}
\usepackage{changes}
\usepackage{graphicx}
\usepackage{color}
\usepackage{gensymb}
\usepackage[numbers,sort&compress]{natbib}
\usepackage{url}
\usepackage{xcolor}
\usepackage{collcell}
\usepackage{threeparttable}
\usepackage{hhline}
\usepackage{pgf}
\usepackage{xstring}
\usepackage{booktabs}
\usepackage{subcaption}
\usepackage{tabularx}
\usepackage{gincltex}
\usepackage{balance}
\usepackage{multirow}
\usepackage{pgfplots}
\usepackage{pgfkeys}
\usepackage{tikzscale}
\usepackage{siunitx}
\usepackage{tikz}
\usepackage[utf8]{inputenc}
\pgfplotsset{compat=newest}
\usepackage{booktabs}
\usetikzlibrary{pgfplots.groupplots}
\definecolor{myred}{rgb}{.8,.0,.0}

\pgfkeys{
  /includetikzgraphic/.is family, /includetikzgraphic,
  default/.style = {
    width = \textwidth,
    ratio = 0.8},
  width/.estore in = \itgWidth,
  ratio/.estore in = \itgRatio
}
\newcommand\blfootnote[1]{%
  \begingroup
  \renewcommand\thefootnote{}\footnote{#1}%
  \addtocounter{footnote}{-1}%
  \endgroup
}
\newlength\figureheight
\newlength\figurewidth

\begin{document}

\bstctlcite{IEEEexample:BSTcontrol}

\title{ Intensity augmentation for domain transfer of whole breast segmentation in MRI  }

\author{L.S. Hesse, G. Kuling, M. Veta, A.L. Martel }

\maketitle

\blfootnote{We acknowledge the support of the Natural Sciences and Engineering Research Council of Canada (NSERC). 

L.S. Hesse is with the department of Biomedical Engineering, Eindhoven University of Technology and with the department of Medical Biophysics, University of Toronto (email: l.s.hesse@student.tue.nl).

G. Kuling is with the department of Medical Biophysics, University of Toronto (email: grey.kuling@sri.utoronto.ca).

M. Veta is with the department of Biomedical Engineering, Eindhoven University of Technology (email: m.veta@tue.nl).

A.L. Martel is with the department of Medical Biophysics, University of Toronto and with Sunnybrook Research Institute, Toronto (email: anne.martel@sri.utoronto.ca).}

\begin{abstract} 

The segmentation of the breast from the chest wall is an important first step in the analysis of breast magnetic resonance images. 3D U-nets have been shown to obtain high segmentation accuracy and appear to generalize well when trained on one scanner type and tested on another scanner, provided that a very similar T1-weighted MR protocol is used. There has, however, been little work addressing the problem of domain adaptation when image intensities or patient orientation differ markedly between the training set and an unseen test set. 
To overcome the domain shift we propose to apply extensive intensity augmentation in addition to geometric augmentation during  training. We explored both style transfer and a novel intensity remapping approach as intensity augmentation strategies.  
For our experiments we trained a 3D U-net on T1-weighted scans and tested on T2-weighted scans. By applying intensity augmentation we increased segmentation performance from a DSC of 0.71 to 0.90. This performance is very close to the baseline performance of training and testing on T2-weighted scans (0.92). Furthermore, we applied our network to an independent test set made up of publicly available scans acquired using a T1-weighted TWIST sequence and a different coil configuration. On this dataset we obtained a performance of 0.89, close to the inter-observer variability of the ground truth segmentations (0.92). 
Our results show that using intensity augmentation in addition to geometric augmentation is a suitable method to overcome the intensity domain shift and we expect it to be useful for a wide range of segmentation tasks.  

\end{abstract}

\begin{IEEEkeywords}
Convolutional neural networks, domain transfer, magnetic resonance imaging, whole breast segmentation 

\end{IEEEkeywords}

\section{Introduction}

\IEEEPARstart{B}{reast} cancer is worldwide the most common cause of death from cancer in woman~\cite{ferlay2019estimating}. Widespread mammography screening programs have been implemented in order to detect breast cancer at an early stage, and since MRI has superior sensitivity to mammography, it is increasingly used for high risk screening~\cite{Warner2011,chiarelli2014effectiveness,kriege2004efficacy}. In addition to detecting and diagnosing breast cancer, breast MR images can also provide valuable information about breast composition. Both the ratio of fibroglandular to fat tissue (\%FGT), and background parenchymal enhancement (\%BPE), defined as the percentage of fibroglandular tissue which enhances after contrast administration, are associated with breast cancer risk~\cite{king2011background, Thompson2019association} and have been used to assess the efficacy of therapeutic interventions~\cite{Wu2015oophorectomy}. There is also the potential for MRI to be used to predict the response of the breast to chemopreventative agents~\cite{Pike2013}.

The first step in the analysis of breast MR images is to obtain a a breast segmentation that separates the breast from the background and chest wall. However, because breast MR scans are 3D volumes, manually segmenting the breast in each slice is very time-consuming and not feasible to implement for large scale screening. For this reason, automated whole breast segmentation methods are essential.

In recent years deep learning methods have been applied widely for segmentation in the medical domain~\cite{litjens2017survey}. In previous work in our group we applied a U-net architecture to segment breasts in MRI scans~\cite{fashandi2019investigation}. We obtained a high accuracy on our own dataset but the performance deteriorated when we applied the trained network to a new dataset. The underlying assumption of neural networks is that train and test data originate from the same data distribution. However, in clinical practice this assumption typically does not hold for different datasets. For this reason, models trained on one image domain often do not perform well on a dataset from a different image domain. The difference in data distribution between two domains is typically called domain shift. In MRI this problem is especially pronounced because the pixel values in a conventional MRI scan are not directly related to a physical quantity. Scans acquired in different clinics or with varying scan protocols can therefore vary substantially in appearance. In a recent study~\citet{zhang2019automatic} showed that a trained U-net can obtain similar performance across different MR scanners, showing that a U-net is able to cope with these scan variations. However, in their study all images were acquired in the axial plane using  non contrast T1-weighted images without fat suppression whereas the domain shift arising from different scan protocols or orientations can be considerably more pronounced. In order to develop MRI segmentation methods which can be used clinically across multiple scanning protocols it is essential to overcome the problem arising from this domain shift.  
\begin{figure*}
\centering
\includegraphics[width=1\textwidth]{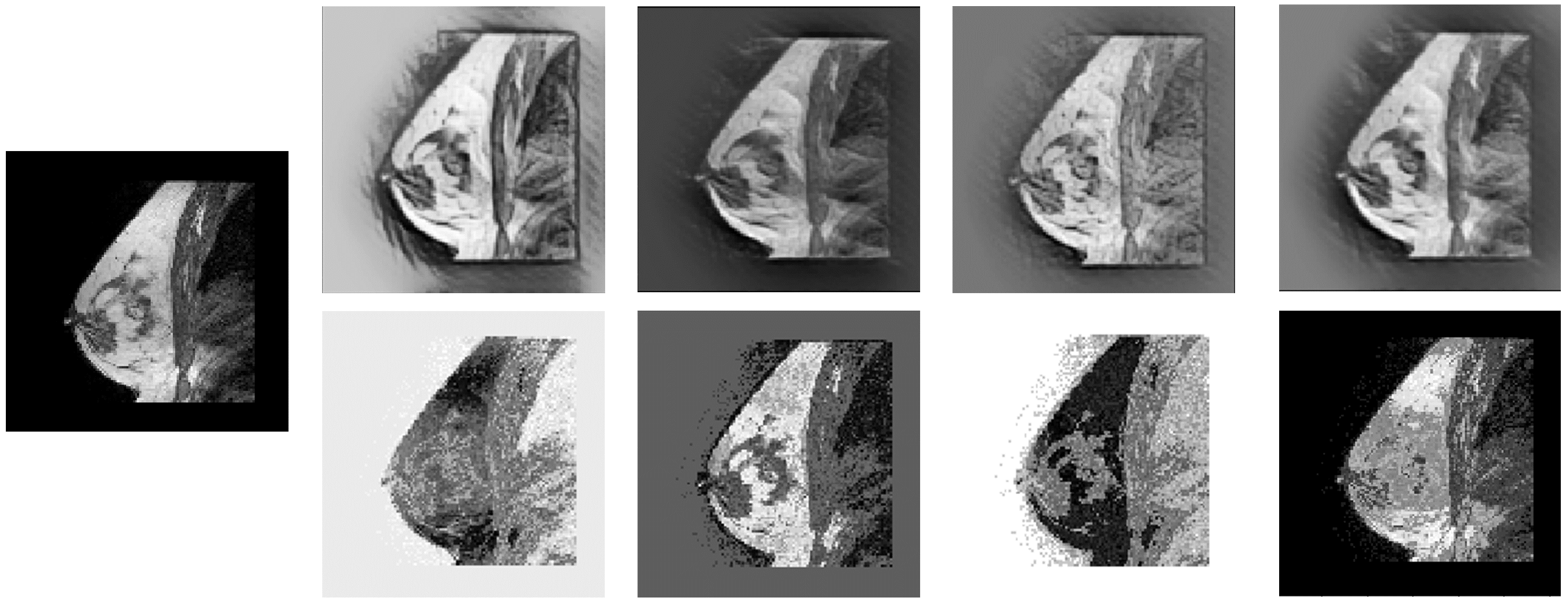}
\caption{Examples of the intensity augmentation methods. The image at the left is the original scan, the top row shows the style augmented scans and the bottom row the intensity remapped scans.}
\label{fig:examplesIntensity}
\end{figure*}

Recent studies have addressed the domain shift problem with varying solutions such as adversarial training or transfer learning~\cite{goodfellow2014generative, csurka2017domain, ghafoorian2017transfer, tajbakhsh2016convolutional}. These methods are typically designed to adapt the model to a specific new domain however this assumes that examples from the new domain are available during training. This may not be true in practice, particularly in medical imaging where acquisition protocols may vary significantly between sites and may change over time. 

One method for generalizing the learned network features to all kind of new image domains is data augmentation~\cite{krizhevsky2012imagenet}. Data augmentation is the artificial transformation of images in order to enlarge the training set and as such prevent overfitting. Traditional augmentation strategies are aimed at simulating realistic images which are expected to occur in a new dataset, such as rotation, translation and contrast stretching. The type of augmentation which is applied to the training data will affect the degree to which a trained model will generalize to new, unseen data. In order to use augmentation to overcome the domain shift in MR breast segmentations, augmentation strategies are necessary which can overcome not only the geometric differences but also the large differences in intensities between MR scans.

To overcome the intensity domain shift in MRI breast segmentation we propose to apply intensity augmentation strategies which produce non-realistic looking MR scans while preserving the image shapes. We expect that by considerably changing the appearance of the input images, the network will learn non-domain specific features and thus show an increased performance on a new dataset. Furthermore, we believe that because images are produced with an unrealistic appearance but a preserved shape, the network will learn shape features rather than texture and intensity features~\cite{tobin2017domain, huang2017arbitrary}. In this study we propose two intensity augmentations methods: style transfer and intensity remapping.

Style transfer is a method in which the texture from a style image is combined with a target image while preserving the semantic content of the target image~\cite{gatys2016image}. It thus changes the color, texture and contrast of an image without altering the geometry~\cite{jackson2018style}. Recent studies showed that style transfer can force a network to focus on shape features~\cite{tobin2017domain} and that it can increase classification performance on a new domain~\cite{jackson2018style}. 

Our intensity remapping approach consists of remapping all pixel values to new pixel values using a randomly generated remapping function. A linear component in the remapping curve ensures the preservation of shape from the input image.  

In summary, we propose to apply two intensity augmentation strategies to create augmented training images not resembling actual MR scans. By heavily distorting the intensities but preserving geometric shape in the images we expect that our segmentation network is forced to focus on the breast shape instead of the breast intensities and will therefore show increased performance on a new domain. Figure~\ref{fig:examplesIntensity} shows examples of the two intensity augmentation strategies applied to a breast MR scan from our dataset.

\section{Related Work} \label{relatedwork}
To overcome the problem of domain shift various solutions have been proposed. These solutions can be divided into domain adaptation and data augmentation. In domain adaptation the model is adapted from a source domain to a specific target domain. In data augmentation the source data is augmented in order to generalize the network without aiming for a specific domain. Technically data augmentation and domain adaptation are quite different concepts. However, they both can be applied to address the same problem: making a model from one domain suitable for the same task in another domain. For this reason, we give a brief overview of related literature in both the fields of domain adaptation and data augmentation. In the last paragraph of this section we present the related work in the field of style transfer.

\paragraph{Domain Adaptation}
Domain adaptation methods can be classified into supervised and unsupervised methods. In supervised methods some labeled data from the target domain is available. This labeled data can be used to adapt a trained model to the target domain, for example fine-tuning a neural network \cite{tajbakhsh2016convolutional}. For some tasks the increase in performance using only a few labeled samples from the target domain can be surprisingly high~\citep{ghafoorian2017transfer}. However, the clinical application of this approach is challenging as it requires labeled data from each new domain. 
 
Unsupervised methods on the other hand use only unlabeled images from the target domain. Two promising approaches of the last couple of years are the discriminative adversarial neural networks (DANNs) and the generative adversarial networks (GANs). In a DANN the network is trained to learn domain invariant features by including a domain discriminative part in the network~\cite{csurka2017domain}. This domain discriminative part has an adversarial objective of not being able to distinguish the different domains, thus enforcing the network to learn domain invariant features. Different implementations of this approach can be found in~\cite{ganin2016domain} and~\cite{tzeng2017adversarial}. 

In a GAN the source domain images are transformed to resemble the images from the target domain~\cite{goodfellow2014generative}. A GAN consists of a generative part, where transformed images are generated, and a discriminative part, which tries to distinguish the transformed images from the target images. By training both parts simultaneously, the network can be optimized to generate very accurate transformations. The resulting transformed source images from the trained generator network can subsequently be used to train a neural network suitable for the target domain.

Although both DANNs and GANs do not need any labeled data, there is usually still the need for unlabeled target data. In clinical practice this constraint can be difficult to achieve as during the development of the software not all domains are known or available. The concept of domain adversarial training can also be applied to datasets not shown before, by assuming that by training a network invariant to the features of two domains, the network will also perform better on another new domain~\cite{lafarge2017domain}. However, this approach will only work if the domain shift between the training and testing domains is not much larger than the shift between the training domains. A more elaborate review on domain adaptation for visual applications can be found in~\cite{csurka2017domain}.

\paragraph{Data Augmentation}
Data augmentation has become well known since the application in AlexNet~\citep{krizhevsky2012imagenet}. It has been widely used to artificially enlarge training datasets and prevent overfitting of neural networks especially in the medical domain, where the availability of large datasets is usually relatively low. Classical augmentation strategies which are commonly applied in medical applications include geometric transformations, addition of noise, Gaussian blurring and histogram based methods~\cite{hussain2017differential}. More closely related to the proposed intensity augmentation strategies is color augmentation, which is commonly used in histopathology to simulate variations in color staining~\cite{tellez2019quantifying}. Color augmentations include perturbations in hue, saturation, brightness and contrast of the image~\cite{tellez2019quantifying, liu2017detecting}. Another possible way to augment color intensities is to apply principal component analysis to the RGB pixel values as applied in~\citep{krizhevsky2012imagenet}.        

\begin{figure*}
\centering
\includegraphics[width=1\textwidth]{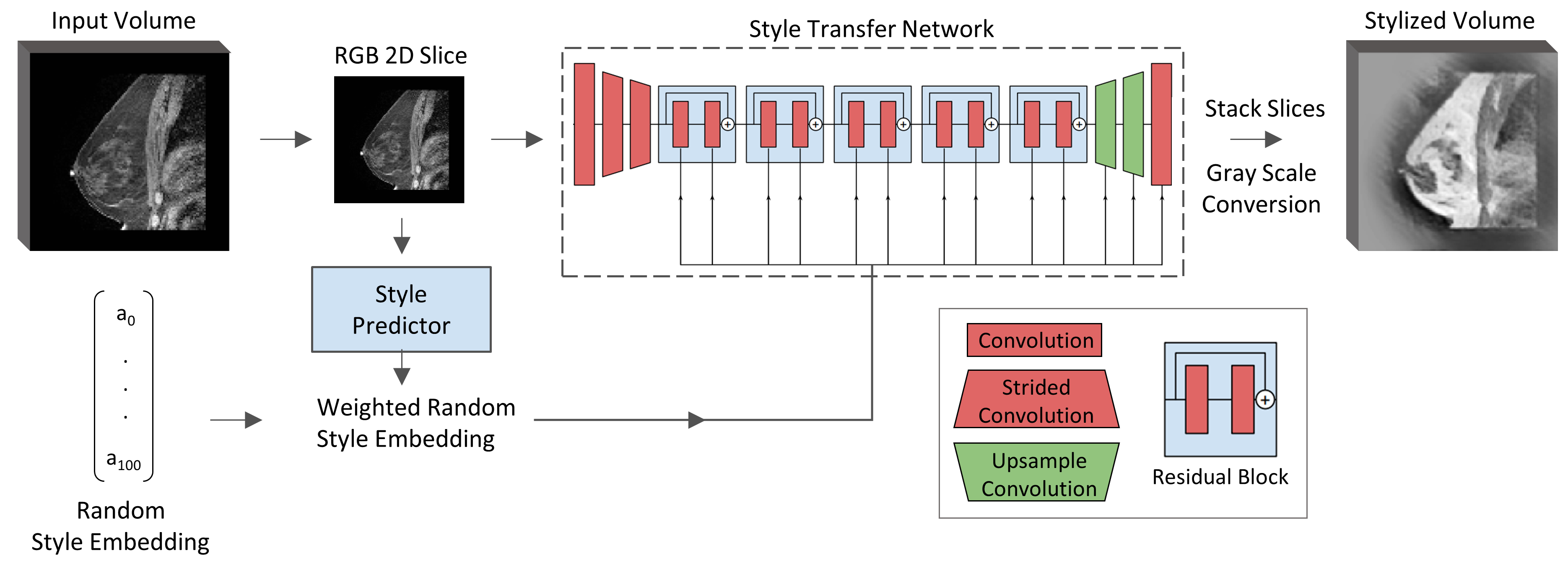}
\caption{Overview of the style transfer pipeline. The input scan volume is converted to 3-channel RGB volume and analyzed per slice. For each scan volume a random style embedding is generated which is combined with the style of the image itself, predicted by the style predictor. The style transfer network converts style and input slice into a stylized image. Adapted from~\cite{jackson2018style}.}
\label{fig:StyleTransfer}
\end{figure*}

\paragraph{Style Transfer}
In this study we propose to use style transfer as intensity augmentation technique. Artistic style transfer using neural networks was first proposed in~\cite{gatys2016image}. Initial style transfer networks were however only trained with a single style, and were not able to apply new styles~\cite{johnson2016perceptual,ulyanov2016texture}. This single style transfer network was subsequently extended to a fully arbitrary style transfer, which is able to generalize to unseen styles~\cite{ghiasi2017exploring, huang2017arbitrary}. Very recently, \citet{geirhos2018imagenet} used the fully arbitrary style transfer of~\cite{huang2017arbitrary} to replace the style of images in the ImageNet dataset with random styles. It was shown that by training a classification network with this new dataset, the network is able to learn a shape-based representation instead of a texture-based one. \citet{jackson2018style} extended the approach proposed in~\cite{ghiasi2017exploring} to augment images with random styles, which where subsequently used to train a classification network. The study showed that style augmentation in combination with regular augmentation was able to increase classifier performance. To the best of our knowledge style augmentation has not yet been applied in either the medical domain or for segmentation tasks.

\section{Methods} \label{methods}

\subsection{Segmentation Network Architecture}
The architecture used in this study is a 3D U-Net, which was first proposed as a 2D U-Net by ~\citet{ronneberger2015u} and extended to a 3D network by~\citet{cciccek20163d}. U-net is a fully convolutional network consisting of an up- and down-sampling path with 'skip' connections to connect high and low level features directly with each other. Every layer block consists of two convolutional layers, each followed by an activation function. In the down-sampling path, also called the analysis path, the down-sampling operations are performed by a max pool layer when proceeding to the next layer block. In the up-sampling path,  also called the synthesis path, an up-sampling layer ensures up-sampling of the output when proceeding to the next layer block. Connections between the analysis and synthesis paths concatenate the feature maps and ensure the transfer of lower level features to the synthesis path. The number of feature maps in the convolutional layers doubles with each down-sampling operation and halves for each up-sampling operation. The rectified linear unit (ReLu) was used as activation function and a sigmoid activation function at the last layer. Dropout layers were added at the beginning of each layer block in the synthesis path.

Based on a grid search in our previous work~\cite{fashandi2019investigation}, we used a network with depth 4, a dropout ratio of 0.2 and stochastic gradient descent (SGD) with a learning rate of 0.01 as an optimizer. The Dice similarity coefficient (DSC) was used as the loss function, which is frequently used for training U-nets~\cite{ronneberger2015u}. The networks were trained and tested on a Nvidia GeForce TITAN X GPU with 12 GB of memory using Tensorflow 1.13.1~\cite{tensorflow2015-whitepaper} and Keras 2.2.4~\cite{chollet2015keras}. Each network was trained for 200 epochs.

\subsection{Augmentation Strategies}
\paragraph{Geometric Augmentation}
Geometric augmentation is applied in order to extend our training set and make the network resistant to geometric difference between datasets. MR breast images from different datasets can vary due to factors such as the field of view of the acquired scan and the used breast coils. We implemented scaling with a factor between 0.8 and 1.2, rotation between -5\degree\ and 5\degree\ and translation with a maximum of 10 mm in plane and 5 mm in the slice direction. The smaller translation in slice direction was selected because of the smaller network input size in this direction. The extremes for the geometric augmentations were determined by inspection of the resulting images, in which we estimated the amount of possible geometric variation to be expected in new datasets.

\begin{figure*}
\centering
\includegraphics[width=0.9\textwidth]{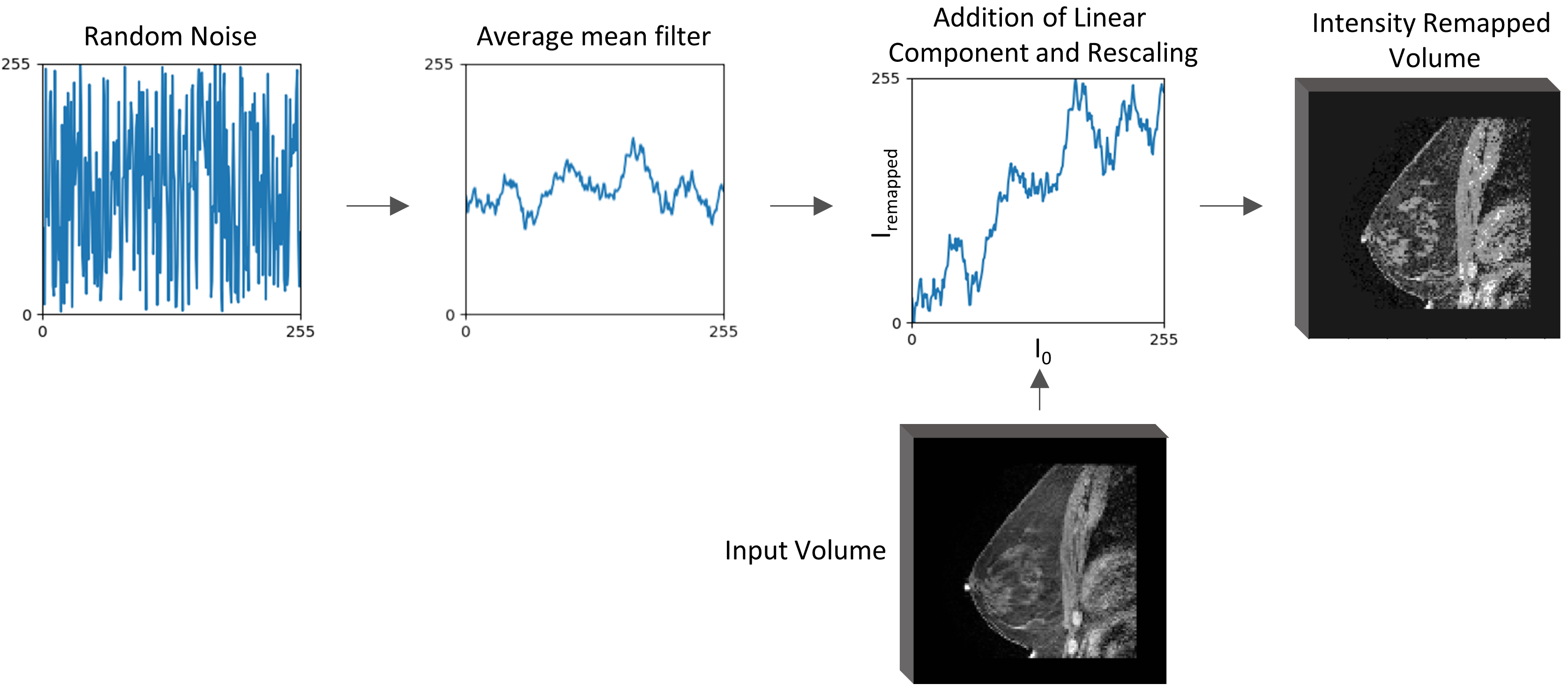}
\caption{Overview of the intensity remapping pipeline. The intensities of the input image are transformed to new intensities using a remapping curve. This remapping curve is generated by filtering a random noise curve with an average mean filter and the addition of a linear component.}
\label{fig:IntensityRemapping}
\end{figure*}

\paragraph{Style Transfer}
We adopted the implementation of~\cite{jackson2018style} for our style transfer. In this approach the style image is replaced by a style embedding $S \in R^{100}$. In order to create random augmentations this style embedding is randomly sampled from a multivariate normal distribution. Furthermore, in order to control the strength of the augmentation, the random style embedding is combined with the style of the image itself resulting in a weighted random style embedding:
\begin{equation}
    S_{weighted} = ( \alpha - 1 )  \cdot S_{random} +  \alpha\cdot  S_{image},
\end{equation}
where $S_{weighted}$ is the weighted style embedding, $S_{random}$ is the random style embedding and $S_{image}$ is the style embedding of the image. The style embedding of the image is extracted using a style predictor network. Both the weighted style embedding and the image are given to the trained style transfer network, resulting in a stylized image. 

Our implementation of this approach is shown in Figure \ref{fig:StyleTransfer}. We used both the trained style predictor network and the trained style transfer network directly from~\citep{jackson2018style}. The input of these networks were 2D RGB channel images. For this reason we converted our gray scale images to a 3-channel input, setting each channel to the same value, and analyzed our 3D volumes slice by slice. For each 3D volume the random style embedding was kept equal for all slices in order to apply the same style to the whole volume. After stylization the slices were stacked and converted to gray scale using:

\begin{equation}
    I = \frac{1}{1000}(299\cdot R+587\cdot G+114\cdot B)
\end{equation}

with I the gray scale intensity and R, G and B the individual channels of the 3-channel output image. The weight of the image style ($\alpha$) was set at 0.5, which was shown to give the best results in~\cite{jackson2018style}.

\paragraph{Intensity Remapping}
The approach we implemented for the intensity remapping is depicted in Figure~\ref{fig:IntensityRemapping}. The pixel values of the original image are replaced by new intensities using an intensity remapping curve. This remapping curve is created separately for each input volume, resulting in randomization of the augmentation.   

In order to construct the remapping curve, random noise is sampled from a uniform distribution over the interval 0-255. This random noise curve is subsequently smoothed with a moving average filter. To preserve some of the intensity relations of the original image, a linear component was then added to the remapping curve. For each remapping curve the linear component was randomly selected to be positive or negative. Finally, the curve was scaled between 0 and 255 to match the pixel values of the input image. For the experiments in this paper we used a window size of 20 for the moving average filter and a linear component size of 0.5. Both parameters were determined by a visual inspection of the remapped images on the amount of desired augmentation.

\paragraph{Implementation}
We implemented the geometric augmentations online in which each scan volume was augmented using SciPy 1.2.1~\cite{scipy}. The augmentation parameters were randomly determined in the predefined ranges. Style randomization was too computationally expensive to perform during training. Therefore we decided to generate all intensity augmented images beforehand. Of each scan volume in the training set, 2 style and 2 intensity remapped volumes were generated. During training of the intensity augmentation experiments a ratio of 1:2, original to augmented respectively, was used. This ratio was determined using preliminary experiments which showed this ratio to give the best results. 

Style randomization was performed on a Nvidia GeForce TITAN X GPU in order to increase speed. Stylization of 1 scan took on average 25 seconds on the GPU, compared to about 120 seconds on the CPU. Intensity remapping of a complete scan took only 0.6 seconds.

\subsection{Datasets}
For this study we used two datasets: A private in house dataset acquired at Sunnybrook Research Institute, and the Breast-QIN DCE-MRI dataset, which is publicly available on the cancer imaging archive~\cite{QIN-data}. 
The Sunnybrook dataset consists of bilateral MR scans acquired on a 1.5T scanner (Signa, General Electric Medical Systems, Milwaukee, WS) using a breast coil. The use of this image data was approved by the institutional review board of Sunnybrook Health Sciences Centre. For each scan volume 4 different scan types are available, which were acquired directly after each other:

\begin{itemize}
    \item T1W sagittal FS: Precontrast T1 weighted (T1W) with fat suppression acquired in sagittal orientation
    \item T1W sagittal WOFS: Precontrast T1W without fat suppression acquired in sagittal orientation
    \item T1W axial FS: Precontrast T1W with fat suppression acquired in axial orientation
    \item T2W sagittal FS: Precontrast T2 weighted (T2W) acquired in sagittal orientation
\end{itemize}

The average resolution of the volumes is 0.38 mm in plane and 3 mm in between slices for the sagittal oriented scans, and 0.66 mm in plane and 1.5 mm between slices for the axial scans.  

The QIN-Breast DCE-MRI dataset contains dynamic contrast-enhanced (DCE) MR scans, acquired with a Siemens 3T TIM Trio system using a four-channel bilateral phased-array breast coil. The axial bilateral DCE-MR scans were acquired with fat suppression with a 3D gradient echo-based time-resolved angiography with stochastic trajectories (TWIST) sequence. The scans have an average resolution of 1 mm in plane 1.4 mm between slices. For each patient, DCE series of two visits are available: prior to treatment and after the first cycle of treatment. We chose to use the DCE series from the first visit and selected the scan acquired at the first time point of the DCE series as this was considered to be closest to the precontrast scans of the Sunnybrook dataset. We excluded one of the scan volumes because it contained breast implants.

The Sunnybrook scan volumes were divided in a training, validation and testing subset. For our experiments we considered inside each subset the 3 T1W scan types as the T1 training/validation/testing subset, and the T2W sagittal FS type as the T2 training/validation/testing subset. The T1 subsets thus contain three times the number of scan volumes indicated, as there are 3 types available for each volume. The QIN-Breast dataset is used as a completely unseen test set and was only used for final testing. The number of whole breast volumes in each dataset can be found in Table~\ref{table:datasets}.

\begin{table}\centering
\begin{threeparttable} 
\caption{Number of breast volumes in the subsets used for training, validating and testing.}
\begin{tabular*}{\linewidth}{@{\extracolsep{\fill}}lcc} \toprule
Data set & Breast Volumes &  Observers \\
\midrule
SB Training data set & 61 & 1   \\ 
SB Validation data set & 8 & 1 \\
SB Test data set  & 15 &  3       \\ \midrule
QIN-Breast & 9 & 1\tnote{*} \\ \bottomrule
\end{tabular*}
\begin{tablenotes}\footnotesize
\item[*] 2 out of 9 volumes were segmented by 3 readers
\end{tablenotes}
\label{table:datasets}
\end{threeparttable}
\end{table}

 \begin{figure*}
\centering
\includegraphics[width=1\textwidth]{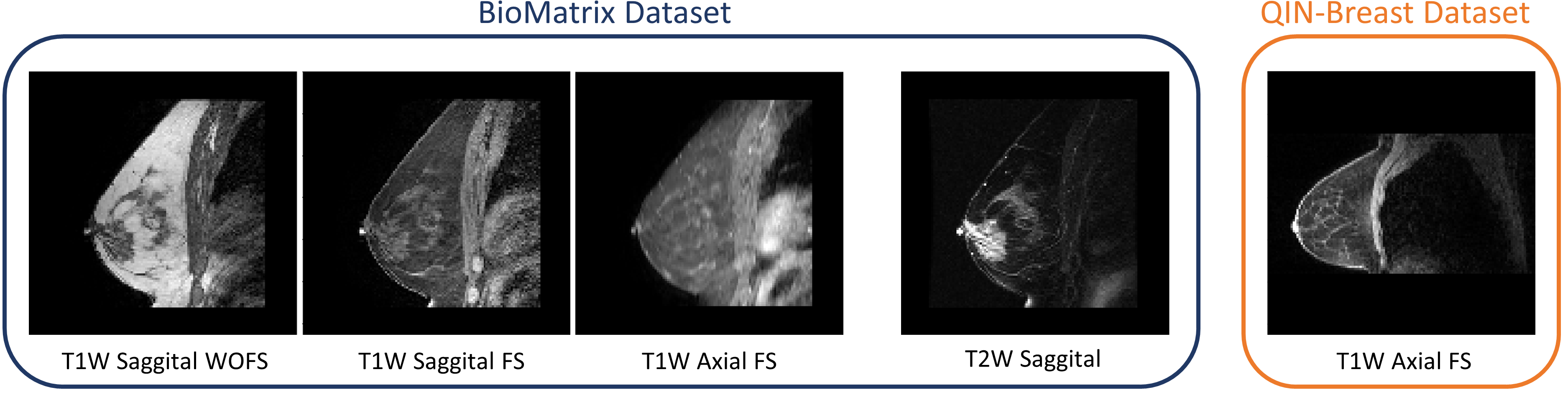}
\caption{Example slices of the two datasets. The images shown were resampled to the axial direction and to the same size and spacing.}
\label{fig:classOverview}
\end{figure*}

\begin{figure}
\centering
\includegraphics[width=0.40\textwidth]{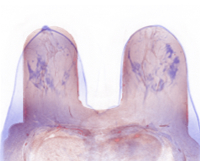}
\caption{Overlay of two breast volumes of similar size from the QIN-Breast (blue) and Sunnybrook (red) dataset. It can be observed that the breast in blue is more relaxed and therefore has a wider shape.}
\label{fig:QinSunOverleay}
\end{figure}

\subsection{Preprocessing} \label{preprocessing}
The whole breast volumes were divided into left and right breast and resampled to the axial orientation. Furthermore, the volumes were resampled to an isotropic resolution of 2 mm and resized to  $64\times 128 \times 128$ pixels using either cropping or padding. The breast volumes in the QIN-Breast dataset were acquired with a different type of coil in which less compression was applied than for the Sunnybrook scans. Because of this reduced compression, the breast volumes in the QIN-Breast dataset were not able to fit completely into the field of view of the network. To illustrate this geometric difference, an overlay of two breasts of similar size from the two datasets is shown in Figure \ref{fig:QinSunOverleay}. To overcome this problem, the QIN-Breast scans were scaled down with a factor of 0.8 before resizing.

\subsection{Ground Truth Generation}
Ground truth segmentations for the Sunnybrook dataset were available from our previous work~\cite{fashandi2019investigation}. In this study we generated random forest segmentations for each T1W sagittal WOFS scan volume as described in~\cite{martel2016breast}. These segmentations were then manually corrected in ITK-SNAP~\cite{yushkevich2006user} and resampled and resized to the same resolution and size as the scan volumes itself. Next, disconnected regions from the 2D slices were removed and the fat tissue under the breast was cut off. Cutting the fat tissue was performed by automatically selecting the most concave point under the breast in each slice and drawing a straight line to the right from this point below which the segmentation was removed. Finally, the lateral boundaries of the breast volume were determined by assuming that the slice where there was most increase in breast volume was the boundary of the breast. The increase in breast volume was calculated by taking the derivative of the size of the mask across slices. The scan volumes in the Sunnybrook test set where each segmented by three readers. These segmentations where combined using the STAPLE algorithm to create a final mask~\cite{warfield2004simultaneous}. 

The ground truth segmentations for the Breast-QIN dataset were generated using ITK-SNAP. In order to automatically segment the breast-air boundary, the segmentation tool of ITK-SNAP was used. Next, every 3\textsuperscript{th}-5\textsuperscript{th} slice in the axial orientation was segmented completely, correcting the initial segmentations and removing the segmentations on intermediate slices. Subsequently, the interpolation tool of ITK-SNAP was used to interpolate the segmentations to the intermediate slices. After interpolation the masks were inspected and manually corrected if interpolation had failed in some slices. The resulting segmentations were scaled down with a factor of 0.8, and resampled and resized to the same resolution and size as the scan volumes. As for the Sunnybrook segmentations, the disconnected regions were then removed from the 2D slices. However, the algorithms which were developed to determine the cutting points and the lateral boundaries of the breast for the Sunnybrook dataset did not work correctly for the Breast-QIN dataset. Because of the different degree of breast compression, as illustrated in Figure \ref{fig:QinSunOverleay}, the lateral boundaries were not as well defined and could not be determined by the largest increase in breast volume. Because of this difference, the concave point under the breast could also not be determined automatically. For this reason we manually selected the cutting points for each slice in the masks, and cut the fat under the breasts based on this selection. Furthermore, the lateral boundaries were determined as the last slice where there was still breast visible. 
Because the QIN-Breast scans had a high resolution, segmenting the volumes was a very time consuming task. For this reason the ten scans were divided between three observers. In order to obtain a measure of inter-observer variability, two scans were segmented by all three observers. The final masks for these two cases were generated using the STAPLE algorithm~\cite{warfield2004simultaneous}. 

\begin{figure*}
	\centering
	\setlength\figureheight{7 cm}
	\setlength\figurewidth{1\textwidth}
	\includegraphics{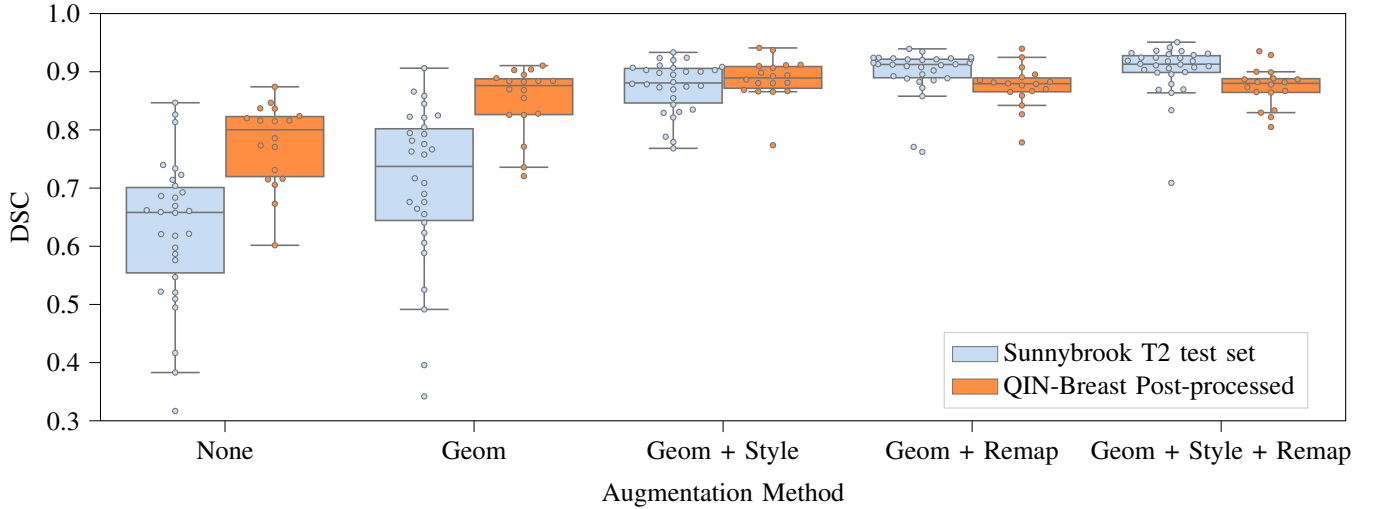}
	\caption{Box plots of the DSC of the different augmentations methods on the Sunnybrook T2 test set and the QIN-Breast dataset after post-processing.}
\label{fig:BoxplotsDice}
\end{figure*}

\subsection{QIN-Breast Post-processing} \label{postprocessing}
Initial resulting segmentations on the QIN-Breast dataset showed segmented regions in the thorax which were not connected to the breast segmentation. In order to focus on the performance of the segmentation of the breast itself, we decided to implement post-processing for this dataset to remove those regions from the segmentations. For each 2D slice we did a connected component analysis and selected the component which was centered most left sided and having a volume larger than 100 pixels. If no component larger than 100 pixels was present, the largest available component was chosen. In some slices after the lateral boundary of the breast, there was still thorax segmentation present after the 2D analysis. In order to remove these volumes as well, we subsequently did a 3D connected component analysis on the whole volume and only kept the largest component.

\section{Experiments and Results}
For the QIN-Breast dataset we created manual ground truth segmentations in this study. To determine the amount of inter-observer variability of our segmentations, two of the ten scans were segmented by all three observers. For each segmentation of a single observer we calculated the average DSC score of the other two observers, which is shown in Table~\ref{DSCVariability}.

In order to obtain a baseline performance for both data types in the Sunnybrook dataset, we first trained and validated our network without any augmentation on the Sunnybrook T1 and T2 data separately. The results of these baseline experiments are shown in Table~\ref{table:baseline}. The baseline performance is a DSC of 0.94 for the T1 scan volumes and 0.92 for the T2 scan volumes. Furthermore, it can be observed that the largest decrease in performance occurs for the network trained with T1 and tested on T2. For this reason we selected T1 images for the training set and T2 images for the validation set in our experiments. 

We trained our network on the T1 training set applying different augmentation strategies. For each augmentation strategy the network was trained five times. The networks were evaluated on the T2 validation set and the network with the highest DSC score on this set was selected. This network was subsequently applied to the T1 and T2 testing set as well as to the scans in the QIN-Breast dataset. The resulting average DSC for each of the augmentation strategies on the different test sets are shown in Table~\ref{table:performance}. For the evaluation on the QIN-Breast datasets, average DSC are shown before and after post-processing of the resulting segmentations. In Figure \ref{fig:BoxplotsDice} box plots of the DSC on the Sunnybrook T2 testing subset and on the QIN-Breast dataset after post-processing are shown. The highest average DSC obtained for the T2 scan volumes is 0.90 which is close to the baseline performance of 0.92. The average DSC of the T1 test set is equal to baseline performance for all augmentation strategies. All three experiments applying intensity augmentation (geometric \& style, geometric \& intensity remapping and geometric \& style \& intensity remapping) increased the performance by a considerable amount. We performed a Friedman test followed by Dunn's multiple comparison test applying Bonferroni P-value correction to correct for the multiple comparisons to statistically analyse these results. We found that on the T2 test subset all intensity augmentation strategies significantly increased performance compared to applying no augmentation or applying only geometric augmentation. A complete overview of the statistical analysis can be found in Appendix~\ref{appendix:exclusion}.

\begin{figure*}
\centering
\includegraphics[width=1\textwidth]{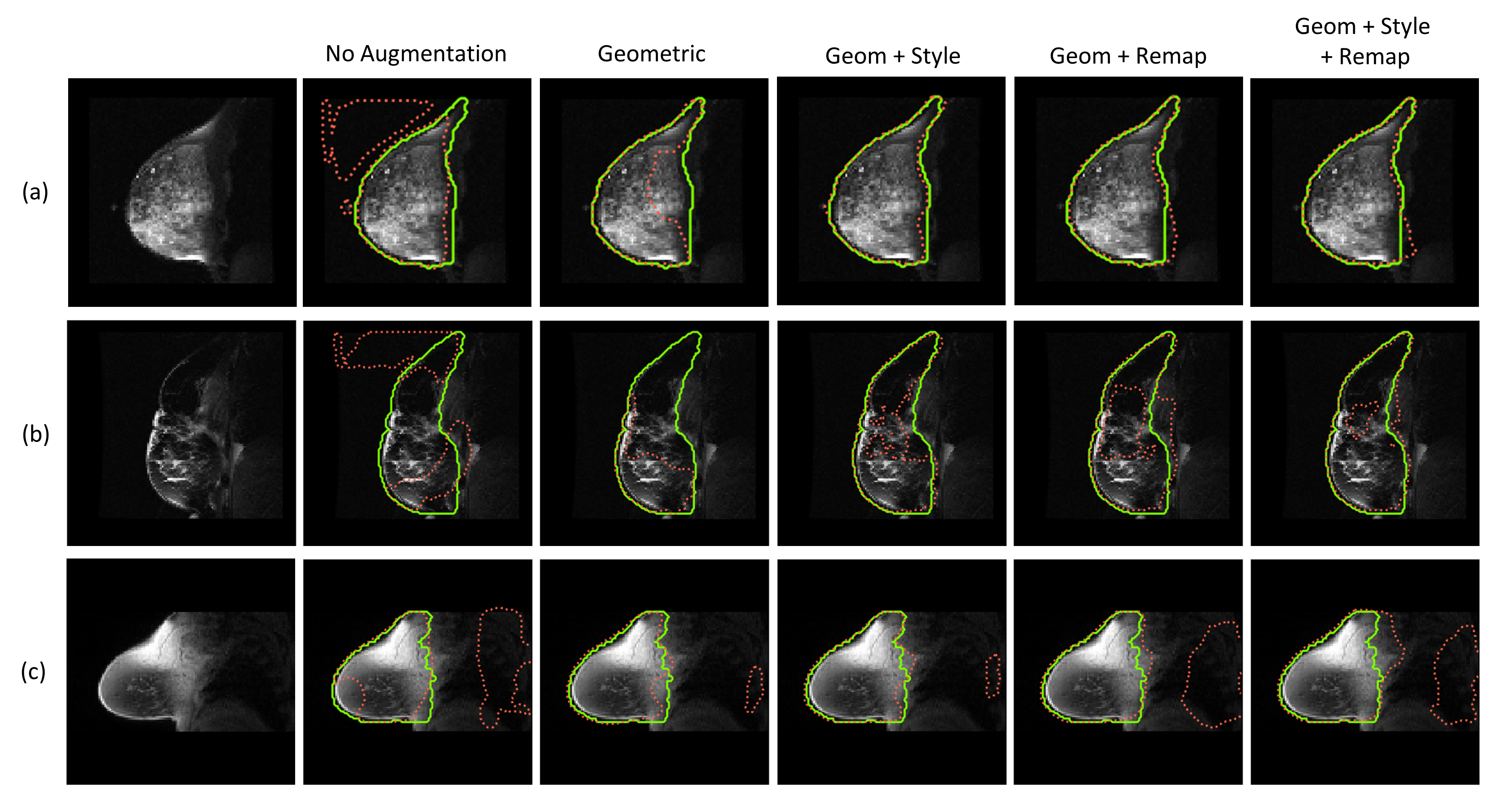}
\caption{Sample output segmentations with different augmentation strategies. The green solid line represent the ground truth segmentation and the dashed red line the predicted segmentation. (a) T2 Test set: high DSC, (b) T2 Test set: low DSC, (c) QIN-Breast: average DSC}
\label{fig:samplesegs}
\end{figure*}

The post-processing of the QIN-Breast dataset increases the performance especially for no augmentation and intensity remapping. The largest increase in performance for this dataset is obtained by the geometric augmentation, which is significant compared to no augmentation. After post-processing all intensity augmentation strategies show a higher average DSC than applying only geometric augmentation, however this difference is not statistically different. The highest DSC is obtained for applying geometric and style augmentation, resulting in a DSC of 0.89. Sample output segmentations for the T2 test set and the QIN-Breast dataset are shown in Figure~\ref{fig:samplesegs}.

\begin{table} \centering
\begin{threeparttable} 
\caption{Baseline experiments for T1 and T2 Sunnybrook test sets.}
\begin{tabular*}{\linewidth}{@{\extracolsep{\fill}}lcc} \toprule
& \multicolumn{2}{c}{Test set} \\
\cmidrule{2-3} 
  & \multicolumn{1}{c}{T1}& 
\multicolumn{1}{c}{T2}  \\ 
\midrule
Trained and Validated on T1 & 0.94 & 0.55   \\
Trained and Validated on T2 & 0.86 & 0.92   \\
\bottomrule
\end{tabular*}
\label{table:baseline}
\end{threeparttable}
\end{table}

\begin{table}\centering
\begin{threeparttable}
\caption{Average DSC score of segmentations by different observers.}
\begin{tabular*}{\linewidth}{@{\extracolsep{\fill}}lcc@{}} \toprule
 & \multicolumn{2}{c}{Average Dice} \\
\cmidrule{2-3} 
Ground Truth & \multicolumn{1}{c}{Scan 1}& 
\multicolumn{1}{c}{Scan 2 }  \\ 
\midrule
Observer 1 & 0.95 & 0.91   \\
Observer 2 & 0.95 & 0.91   \\
Observer 3 & 0.94 & 0.88 \\
\bottomrule
\end{tabular*}
\label{DSCVariability}
\end{threeparttable}
\end{table}

\begin{table*}\centering
\begin{threeparttable}
\caption{Resulting average DSC using different augmentation methods. The values between brackets are the standard deviations.   }
\begin{tabular*}{\linewidth}{@{\extracolsep{\fill}}lcccc@{}} \toprule
 & \multicolumn{2}{c}{Sunnybrook test set}  & \multicolumn{2}{c}{QIN-Breast}   \\
\cmidrule{2-3}  \cmidrule{4-5}
Augmentation Method & T1 & T2 & before post-processing  & post-processed  \\ \midrule
None & 0.88 (0.04)  & 0.63 (0.12) & 0.61 (0.14) & 0.78 (0.07)\\
Geometric & 0.94 (0.03)  & 0.71 (0.14) & 0.83 (0.06) & 0.85 (0.06)        \\ 
Geometric + Style & 0.94 (0.03) & 0.87 (0.04)&0.85 (0.08) & 0.89 (0.04) \\
Geometric + Remapping & 0.94 (0.02) & 0.90 (0.04) & 0.77 (0.09) & 0.88 (0.04)     \\

Geometric + Remapping + Style & 0.94 (0.02) & 0.90 (0.04) & 0.76 (0.10) & 0.87 (0.03) \\
\bottomrule
\end{tabular*}
\label{table:performance}
\end{threeparttable}
\end{table*}

\section{Discussion and Conclusion} \label{discussion}
Our results show that intensity augmentation can increase the performance of whole breast segmentation on a new domain without decreasing the performance on the training domain. This confirms our hypothesis that by heavily disturbing the intensities in the image while preserving shape, the network is forced to focus on non-domain specific features during training. Furthermore, our findings are in line with recent literature applying style transfer in classification tasks ~\cite{geirhos2018imagenet, jackson2018style}, and show that style transfer, as well as intensity remapping, can also be used for the domain transfer of a segmentation task. 

The intensity domain shift is best demonstrated by the performance on the Sunnybrook T2 test set. For these experiments the network is trained on T1W scans and tested on T2W scans from the same dataset. Therefore the intensity domain shift can be analyzed separately from a geometric domain shift. For this test set we achieved an increase in performance from a DSC of 0.62 to 0.90 by applying intensity remapping and geometric augmentation. The same average performance was achieved when combining aforementioned methods with style transfer. This performance is close to the T2 baseline performance of 0.92, showing that the domain shift between T1W and T2W scans can be almost completely overcome by applying intensity augmentation. Furthermore, for all augmentation experiments the T1W performance remains at 0.94 which indicates that the performance on the training domain is not decreased by the intensity augmentation.

The QIN-Breast dataset was used only for testing, and gives therefore an indication of real-life clinical performance in which the test set is not available beforehand. The scans in this dataset are axial T1W FS scans. This scan type is also present in the Sunnybrook training set which makes the intensity differences smaller than for the T2W scans. The geometric differences between the Sunnybrook and the QIN-Breast scans are however very pronounced. These differences arise mainly from the use of a different breast coil and the larger part of the thorax in the field of view. The largest increase in performance for the QIN-Breast dataset is obtained by applying geometric augmentation. This corresponds to the observation that the geometric domain shift for this dataset is more pronounced than the intensity shift.

Before post-processing of the resulting segmentations of the QIN-Breast scans, intensity remapping significantly decreased the performance compared to applying only geometric augmentation. We observed that this decrease in performance resulted from large segmented areas in the thorax which were less prominant for the geometric and style augmentation. An explanation for this could be that due to the intensity remapping, shapes in the thorax are recognized as possible breast tissue. Because we only applied our best model from the T2 validation subset on our QIN-Breast dataset we can not conclude whether this is a characteristic of the intensity remapping or due to variation in the trained models. 

In order to still be able to obtain a quantitative measure of how well the breast shape itself was segmented, we implemented post-processing to remove the unconnected thorax segmentations. After post-processing, the intensity remapping performed equally well as the other augmentation methods, showing that it is able to correctly distinguish the breast shape in most cases. We achieved an average DSC of 0.89 by applying both geometric and style augmentation. This performance is close to the average inter-observer variability we obtained (DSC of 0.92), showing that our method achieves good performance on a completely new dataset.     

The performance on the T2 Sunnybrook test set showed clearly that there is a significant increase in performance by applying our proposed intensity augmentation techniques. However, it is not apparent from our experiments whether style or intensity remapping performs best. In addition to performance, there are some key differences between the methods which should be considered. The disadvantage of style transfer is that it is essentially a 2D method which is implemented for each slice, potentially resulting in discontinuities in the intensities. Furthermore, the style transfer for the 3D volumes had to be executed on a GPU in order to perform on reasonable speed. Intensity remapping on the other hand can be implemented directly in 3D and is considerably faster. However, we observed that parameter tuning was more challenging for the intensity remapping, as the appearance of the resulting images varied greatly depending on the parameter choice. For the style transfer only the strength of the style augmentation had to be controlled which was relatively straightforward to determine. 

In this study we did not do an elaborate parameter optimization for the augmentation parameters. We think that optimizing these parameters could contribute to an even higher performance. It would be interesting to see whether using a few scans from the Breast-QIN dataset for parameter optimization and model selection increases performance. This could potentially also overcome the necessity for post-processing. Another point of interest are the ground truth segmentations. Whole breast segmentation is currently not carried out in clinical practice and as such there is little consensus about it. In this work we adopted the segmentation guidelines used in~\cite{fashandi2019investigation} but we are intending to create ground truths based on anatomical landmarks for further work.   

The main advantage of our proposed intensity augmentation techniques is that they generalize the network features without aiming for a specific domain. Especially for the clinical implementation of a product this is very advantageous, as it overcomes the need for data from all possible domains. Furthermore, intensity augmentation is relatively easy to implement and can be used in combination with other domain adaptation methods. In this study we applied the intensity augmentation for whole breast segmentations. However, we believe that the proposed augmentation techniques can be applied to a wide range of segmentation tasks.

\section{Acknowledgment}
The authors would like to acknowledge the contribution of Homa Fashandi to the development of software used in this work.

\renewcommand*{\bibfont}{\footnotesize}
\bibliographystyle{IEEEtranN}
\bibliography{bibliography.bib}

\setcounter{figure}{0}  
\setcounter{table}{0}
\clearpage

\onecolumn
\appendices

\section{Statistical results} 
\def\colorModel{gray}

\newcommand\ColCellH[1]{
  \StrDel{#1}{\textless}[\numb]
  \pgfmathparse{\numb<0.05?0:1}  
    \ifnum\pgfmathresult=0\relax\color{white}\fi
    \ifnum\pgfmathresult=0\pgfmathsetmacro\compA{0.5}\fi      
    \ifnum\pgfmathresult=1\relax\color{black}\fi
    \ifnum\pgfmathresult=1\pgfmathsetmacro\compA{0.9}    \fi 
  \edef\x{\noexpand\centering\noexpand\cellcolor[\colorModel]{\compA}}\x #1
  } 
  
\newcolumntype{H}{>{\collectcell\ColCellH}m{1 cm}<{\endcollectcell}}

\newcolumntype{R}[2]{%
    >{\adjustbox{angle=#1,lap=\width+1.3em}\bgroup}%
    l%
    <{\egroup}%
}

\newcommand*\rot[2]{\multicolumn{1}{R{#1}{#2}}}

\begin{table*}[h]\centering
\caption{Statistical analysis on the DSC performance of the different augmentation methods. Results were obtained with a Friedman test followed by Dunn's test with Bonferroni correction for multiple comparisons. Entries are the p-values resulting from the multiple comparison testing. Boxes are colored dark if they are significantly different (p<0.05) from each other and light if there is no statistically significant difference.  }
    \begin{subtable}{1\textwidth}
        \centering
        \begin{tabular}{r*{5}{H}} 
          & \rot{90}{1em}{No Augmentation}& 
        \rot{90}{0em}{Geom} &
        \rot{90}{1em}{Geom + Style}  &
        \rot{90}{1em}{Geom + Remap}&
        \rot{90}{1em}{Geom + Style + Remap} \\ 
        No Augmentation & 1 & 1  & \textless0.01 & \textless0.01 & \textless0.01   \\
        Geom & 1 & 1 & \textless0.01 & \textless0.01 & \textless0.01 \\
        Geom + Style & \textless0.01 & \textless0.01 & 1 & 0.34 & 0.03  \\
        Geom + Remap & \textless0.01& \textless0.01 &0.34 & 1 &1 \\
        Geom + Style + Remap & \textless0.01 & \textless0.01 & 0.03 & 1 &1  \\
        \end{tabular}
        \vspace*{1 mm}
        \subcaption{Sunnybrook T2 test set}
    \end{subtable}
    \begin{subtable}{1\textwidth}
        \centering
        \begin{tabular}{r*{5}{H}} 
          & \rot{90}{1em}{No Augmentation}& 
        \rot{90}{0em}{Geom} &
        \rot{90}{1em}{Geom + Style}  &
        \rot{90}{1em}{Geom + Remap}&
        \rot{90}{1em}{Geom + Style + Remap} \\ 
        No Augmentation & 1& 0.02  & \textless0.01 & \textless0.01 & \textless0.01   \\
        Geom & 0.02 & 1  & 1 & 1 &1 \\
        Geom + Style & \textless0.01 & 1 & 1& 1 & 1  \\
        Geom + Remap & \textless0.01&1 &1 &1 &1 \\
        Geom + Style + Remap & \textless0.01 & 1 & 1 & 1 &1  \\
        \end{tabular}
        \vspace*{1mm}
        \subcaption{QIN-Breast dataset after post-processing}
        \label{table:StatisticsT2}
    \end{subtable}
    \begin{subtable}{1\textwidth}
        \centering
        \begin{tabular}{r*{5}{H}} 
          & \rot{90}{1em}{No Augmentation}& 
        \rot{90}{0em}{Geom} &
        \rot{90}{1em}{Geom + Style}  &
        \rot{90}{1em}{Geom + Remap}&
        \rot{90}{1em}{Geom + Style + Remap} \\ 
        No Augmentation & 1 & \textless0.01  & \textless0.01 & 0.02 & 0.03   \\
        Geom & \textless0.01 & 1  & 1 & 0.20 & 0.15 \\
        Geom + Style & \textless0.01 & 1 & 1& \textless0.01 &\textless0.01  \\
        Geom + Remap & 0.02 & 0.20 &\textless0.01 &1 &1 \\
        Geom + Style + Remap & 0.03 & 0.15 & \textless0.01 & 1 &1  \\
        \end{tabular}
        \vspace*{1mm}
        \subcaption{QIN-Breast dataset before post-processing}
        \label{table:StatisticsT2}
    \end{subtable}
\end{table*}
\label{appendix:exclusion}
\vfill

\end{document}